\documentclass[twocolumn,prc,showpacs,preprintnumbers,amsmath,amssymb,superscriptaddress,nofootinbib]{revtex4-2}

\usepackage{graphicx}
\usepackage{dcolumn}
\usepackage{bm}
\usepackage{booktabs}
\usepackage{placeins}
\usepackage{rotating}
\usepackage{color}

\begin{document}


\title{Investigating the effects of precise mass measurements of Ru and Pd isotopes on machine learning mass modeling}
\author{W. S. Porter}
\email[Corresponding author: ]{wporter@nd.edu}
    \affiliation{Department of Physics and Astronomy, University of Notre Dame, Notre Dame, Indiana 46656, USA}

\author{B. Liu}
    \affiliation{Department of Physics and Astronomy, University of Notre Dame, Notre Dame, Indiana 46656, USA}
    \affiliation{Physics Division, Argonne National Laboratory, Lemont, Illinois 60439, USA}
    
\author{D. Ray}
    \altaffiliation{Current address: TRIUMF, 4004 Wesbrook Mall, Vancouver, British Columbia V6T 2A3, Canada}
    \affiliation{Physics Division, Argonne National Laboratory, Lemont, Illinois 60439, USA}
    \affiliation{Department of Physics and Astronomy, University of Manitoba, Winnipeg, Manitoba R3T 2N2, Canada}

\author{A. A. Valverde}
    \affiliation{Physics Division, Argonne National Laboratory, Lemont, Illinois 60439, USA}
    \affiliation{Department of Physics and Astronomy, University of Manitoba, Winnipeg, Manitoba R3T 2N2, Canada}

\author{M. Li}
    \affiliation{Department of Physics and Astronomy, University of Notre Dame, Notre Dame, Indiana 46656, USA}
    \affiliation{Department of Physics, University of California, Berkeley, Berkeley, California, 94720, USA}

\author{M. R. Mumpower}
    \affiliation{Theoretical Division, Los Alamos National Laboratory, Los Alamos, New Mexico 87545, USA}
    \affiliation{Center for Theoretical Astrophysics, Los Alamos National Laboratory, Los Alamos, New Mexico 87545, USA}
    
\author{M. Brodeur}
    \affiliation{Department of Physics and Astronomy, University of Notre Dame, Notre Dame, Indiana 46656, USA}

\author{D. P. Burdette}
    \affiliation{Physics Division, Argonne National Laboratory, Lemont, Illinois 60439, USA}

\author{N. Callahan}
    \affiliation{Physics Division, Argonne National Laboratory, Lemont, Illinois 60439, USA}
    
\author{A. Cannon}
    \affiliation{Department of Physics and Astronomy, University of Notre Dame, Notre Dame, Indiana 46656, USA}

\author{J. A. Clark}
    \affiliation{Physics Division, Argonne National Laboratory, Lemont, Illinois 60439, USA}
    \affiliation{Department of Physics and Astronomy, University of Manitoba, Winnipeg, Manitoba R3T 2N2, Canada}

\author{D. E. M. Hoff}
    \affiliation{Nuclear and Chemical Sciences Division, Lawrence Livermore National Laboratory, Livermore, California, 94550, USA}
    
\author{A. M. Houff}
    \affiliation{Department of Physics and Astronomy, University of Notre Dame, Notre Dame, Indiana 46656, USA}

\author{F. G. Kondev}
    \affiliation{Physics Division, Argonne National Laboratory, Lemont, Illinois 60439, USA}

\author{A. E. Lovell}
    \affiliation{Theoretical Division, Los Alamos National Laboratory, Los Alamos, New Mexico 87545, USA}

\author{A. T. Mohan}
    \affiliation{Computational Division, Los Alamos National Laboratory, Los Alamos, New Mexico 87545, USA}

\author{G. E. Morgan}
    \affiliation{Physics Division, Argonne National Laboratory, Lemont, Illinois 60439, USA}
    \affiliation{Department of Physics and Astronomy, Louisiana State University, Baton Rouge, Louisiana 70803, USA}
    
\author{C. Quick}
    \affiliation{Department of Physics and Astronomy, University of Notre Dame, Notre Dame, Indiana 46656, USA}
    
\author{G. Savard}
    \affiliation{Physics Division, Argonne National Laboratory, Lemont, Illinois 60439, USA}
    \affiliation{Department of Physics, University of Chicago, Chicago, Illinois 60637, USA}
    
\author{K. S. Sharma}
    \affiliation{Department of Physics and Astronomy, University of Manitoba, Winnipeg, Manitoba R3T 2N2, Canada}

\author{T. M. Sprouse}
    \affiliation{Theoretical Division, Los Alamos National Laboratory, Los Alamos, New Mexico 87545, USA}

\author{L. Varriano}
    \altaffiliation{Current address: Center for Experimental Nuclear Physics and Astrophysics, University of Washington, Seattle, Washington 98195, USA}
    \affiliation{Physics Division, Argonne National Laboratory, Lemont, Illinois 60439, USA}
    \affiliation{Department of Physics, University of Chicago, Chicago, Illinois 60637, USA}


\date{\today}

\begin{abstract}
Atomic masses are a foundational quantity in our understanding of nuclear structure, astrophysics and fundamental symmetries. The long-standing goal of creating a predictive global model for the binding energy of a nucleus remains a significant challenge, however, and prompts the need for precise measurements of atomic masses to serve as anchor points for model developments. We present precise mass measurements of neutron-rich Ru and Pd isotopes performed at the Californium Rare Isotope Breeder Upgrade facility at Argonne National Laboratory using the Canadian Penning Trap mass spectrometer. The masses of $^{108}$Ru, $^{110}$Ru and $^{116}$Pd were measured to a relative mass precision $\delta m/m \approx 10^{-8}$ via the phase-imaging ion-cyclotron-resonance technique, and represent an improvement of approximately an order of magnitude over previous measurements. These mass data were used in conjunction with the physically interpretable machine learning (PIML) model, which uses a mixture density neural network to model mass excesses via a mixture of Gaussian distributions. The effects of our new mass data on a Bayesian-updating of a PIML model are presented.  

\end{abstract}

\maketitle

\section{Introduction}

The mass of an isotope is a critically important nuclear observable for our understanding of the structure of a nucleus
, reactions in stellar environments
, and fundamental interactions 
\cite{Dilling2018}. As such, a predictive model of the atomic mass has been a long-standing aim of the nuclear physics community, with the ultimate goal of determining the masses of exotic nuclei very far from the valley of stability. Conventional mass models are optimized on all available experimental mass measurements \cite{Duflo1995} to find the best set of descriptive parameters, which are used to extrapolate masses of unmeasured nuclei. While most mass models well-describe experimental data, with deviations commonly on the order of 100 keV \cite{Goriely2010,Wang2014,Moller2016} they diverge (on the order of 1 MeV) from each other quickly outside the realm of available data. These deviations motivate the need for further mass measurements of nuclei; both those which remain unmeasured and those that can be measured more precisely and accurately.

While the need for high-precision mass measurements is well established, future developments and innovations are still warranted in the realm of mass models. Given the impossibility of solving the complex, many-body Hamiltonian for all nuclei, models make approximations to allow for computational feasibility. The result is often a fixed model with the focus on the optimization of specific parameters, which provides a limited search into the plethora of potential correlations between nuclei across the nuclear chart. Recently, a different approach has been taken, where the model itself is no longer fixed but rather optimized by using machine learning (ML) techniques \cite{Niu2022,Bobyk2022,Neufcourt2020,Utama2016} with a neural network. This allows for a more wide-reaching, robust search of mass correlations, which could potentially shed light on new physical structure hidden in the unprobed mass surface. Previous work has shown such techniques can well-describe the nuclear mass, both using solely mass excess data \cite{Lovell2022} and when paired with a chosen physical constraint \cite{Mumpower2022,Mumpower2023,Li2024}. While these efforts, alongside a wealth of other ML modeling approaches, have been successful, there remains a wide space of mass correlations to explore with much promise. Similarly to their fixed model counterparts, ML models require precise experimental measurement of nuclear masses to improve their ability to model existing data and test their predictive capabilities.

In this paper, we present precision mass measurements of $^{108}$Ru, $^{110}$Ru and $^{116}$Pd with the Canadian Penning Trap (CPT) using radioactive beams from the Californium Rare Isotope Breeder Upgrade (CARIBU) at Argonne National Laboratory (ANL). All masses were measured to a relative mass precision $\delta m/m \approx 10^{-8}$ using the Phase-Imaging Ion-Cyclotron-Resonance (PI-ICR) technique, and represent an improvement in precision of approximately an order of magnitude over previous results which employed the Time-of-Flight Ion-Cyclotron-Resonance (ToF-ICR) technique \cite{Dilling2018,Hager2007}. These masses were used in the training of the Physically Interpretable Machine Learning (PIML) model, which uses a Mixture Density Network (MDN) approach to determine mass excesses via a admixture of Gaussians.


\section{Experiment}

Ru and Pd mass measurements were performed using the CPT housed at ANL \cite{Savard2001} and coupled to the CARIBU facility \cite{Savard2008}. The radioactive beams of interest were produced by the spontaneous fission of a $^{252}$Cf source of an effective strength of approximately 0.5 Ci. Fission fragments are collected by a large-volume gas catcher \cite{Savard2003}, where they thermalize via collisions with high-purity He gas. Pd and Ru isotopes exit the gas catcher as ions with a charge state of $q = 1$, and are delivered to a high-resolution isobar separator \cite{Davids2008}, which selects a unique $A/q$, where $A$ is the mass number, via two bending magnets.

Isobaric beams progress into the radio-frequency quadrupole (RFQ) cooler-buncher, which cools beams via collisions with a He buffer gas at $\approx 10^{-4}$ torr and bunches beams via collection in a time-varying potential well. Cooled ion bunches, which include the Pd and Ru isotopes of interest (IOIs), are ejected and delivered to the multiple-reflection time-of-flight mass separator (MR-ToF-MS) \cite{Hirsh2016}, which separates ions via their unique time-of-flight over a given path and kinetic energy \cite{Yavor2009}. After undergoing between 580 to 850 isochronous turns between electrostatic mirrors, ions are ejected and a unique time-of-flight range is selected via a Bradbury-Nielson gate. The MR-ToF-MS, with a mass resolving power of approximately $10^5$, delivers highly selective ion bunches of Pd and Ru isotopes to a linear Paul trap for further cooling and accumulation before injection into the Penning trap.

Masses of ions are determined via their cyclotron frequency ($\nu_c$) in a magnetic field. The CPT is a hyperbolic Penning trap consisting of the standard hyperbolic ring and endcap electrodes, with additional correction ring and tube electrodes, all housed within a $\approx$ 6 T magnetic field. Ionic motion inside a Penning trap can be decomposed into three eigenmotions with frequencies $\nu_+$, $\nu_-$ and $\nu_z$. $\nu_c$, to first approximation given as the sum of the radial components $\nu_+$ and $\nu_-$ \cite{Brown1986}, is determined via the PI-ICR technique \cite{Eliseev2013}, where the motional phases of ions within the trap are projected onto a position-sensitive microchannel plate (PS-MCP) detector \cite{Jagutzki2002}.

The cyclotron frequency is determined via a reference phase measurement and a final phase measurement. In both, the IOI is injected into the Penning trap from the linear Paul trap, where they are centered via the application of a dipole radiofrequency (RF) pulse of frequency $\nu_-$ and excited to a radius by the application of a dipole RF pulse of frequency $\nu_+$. In a reference phase measurement, a quadrupole RF pulse at $\nu_c$ is applied immediately after the $\nu_+$ pulse. In a final phase measurement, the $\nu_c$ pulse is delayed by a time $t_{\text{acc}}$, such that the ions acquire a mass-dependent phase during this so-called accumulation time. In both cases, the ions are ejected into a drift tube before detection via a PS-MCP. The cyclotron frequency is given by the difference in phases between these measurements, such that:

\begin{equation}
    \nu_c = \frac{\Delta\phi}{2\pi t_{\text{acc}}} = \frac{\phi_f - \phi_i + 2\pi n(t_{\text{acc}})}{2\pi t_{\text{acc}}}
\end{equation}

\noindent where $n(t_{\text{acc}})$ is the number of full revolutions undergone by the ion in time $t_{\text{acc}}$, $\phi_f$ and $\phi_i$ are the final and reference phases, respectively. A measurement of a well-known mass of the same $A/q$ at high statistics is used to calibrate the magnetic field strength, and the ratio $R = \nu_{c,\text{IOI}}/\nu_{c,\text{cal}}$ between the IOI and the calibrant was used to determine the Pd and Ru masses. More information on the employed measurement scheme can be found in \cite{Orford2020}.

\section{Analysis}

\begin{figure}[tb]
    \begin{center}
        \includegraphics[width=1\columnwidth]{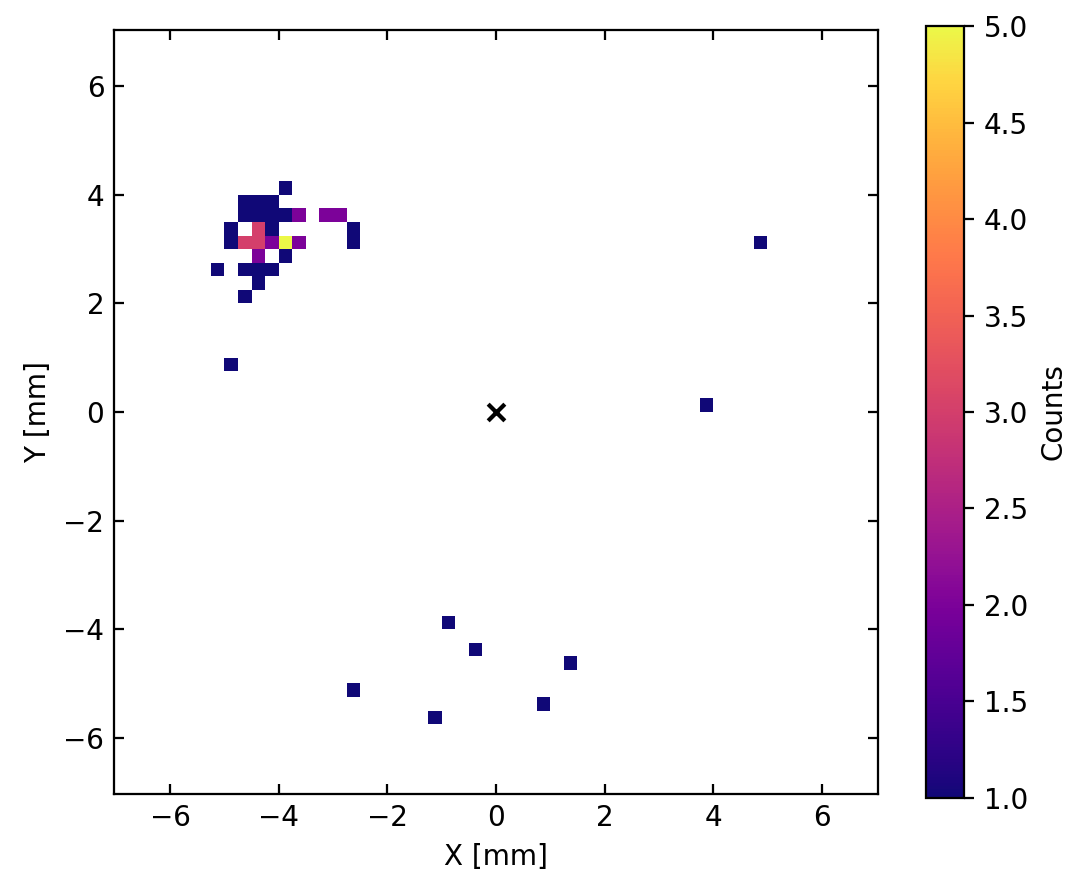}
        \caption{A plot of counts of $^{110}$Ru isotopes detected during a PI-ICR measurement with accumulation time $t_\text{acc} = 450.432$ ms.}
        \label{fig:clusters}
    \end{center}
\end{figure}

In order to determine $\phi_f$ and $\phi_i$, the positional centers of the spots resulting from ion detection at the PS-MCP must be obtained. This is accomplished by fitting Gaussians to all data points assigned to a given spot. Spot assignment is determined via data clustering using the mean shift algorithm \cite{Comaniciu2002}, similar to that of \cite{Orford2020a,Orford2022}. An example of a result of the clustering algorithm is shown in Fig. \ref{fig:clusters}. Two one-dimensional (1D) Gaussians are fit to clusters using a maximum likelihood estimator, and the mean and standard deviation are taken as the spot center and uncertainty in position space.

\begin{figure}[tb]
    \begin{center}
        \includegraphics[width=1\columnwidth]{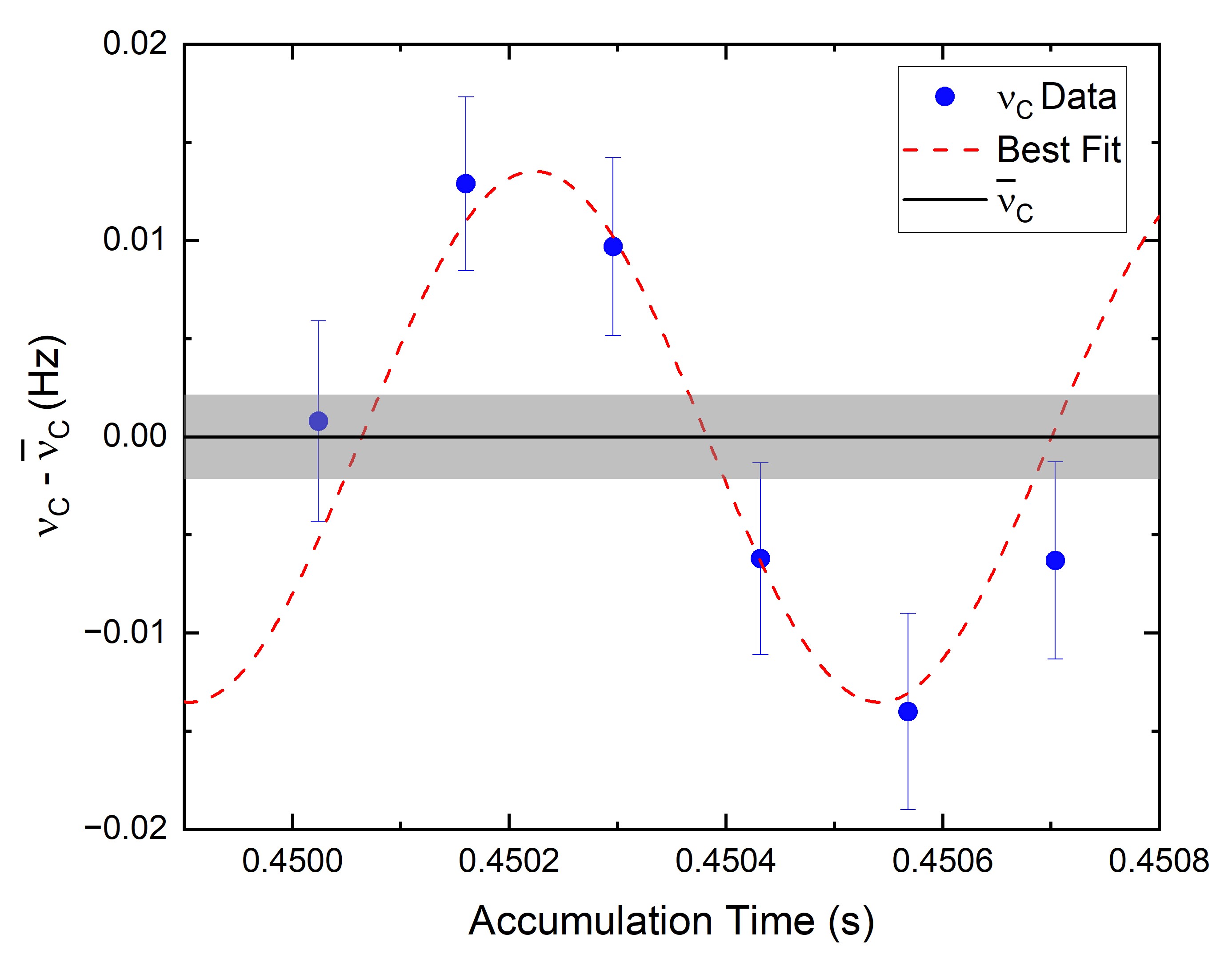}
        \caption{A plot of measured $\nu_c$ values for $^{110}$Ru at various accumulation times spanning 450.024 and 450.704 ms. Blue circles represent $\nu_c$ values, and the red dashes represent the best fit sine curve. $\overline{\rm \nu_c}$ and the associated 1$\sigma$ uncertainty are shown by the black line and surrounding grey band.} 
        \label{fig:sine}
    \end{center}
\end{figure}

A number of systematic effects must be accounted for during the data-taking and analysis process. First, the presence of initial magnetron motion of the ions upon injection into the Penning trap results in an extra oscillatory component after excitation to the revolution radius. This manifests as a sinusoidal dependence of $\phi_f$, and subsequently the cyclotron frequency, on the accumulation time, with a period determined by the trap's characteristic $\nu_-$. To fully characterize this motion, data is taken at various $t_{\text{acc}}$ within a $\approx$ 1 ms window to encompass a period of magnetron motion, and a three parameter sinusoidal model, described in \cite{Orford2020}, is fitted to extract the true cyclotron frequency ($\overline{\rm \nu_c}$). An example of a fitted sinusoid and the associated cyclotron frequency data is presented in Fig. \ref{fig:sine}.

Second, the presence of multiple species in the trap results in the reference phase being an amalgamation of the slightly different phases acquired by each ionic species during excitation. To properly determine the phase difference for a given isotope, the reference phase is shifted to account for the different populations of species present in a spectrum. This process is described in greater detail in \cite{Orford2020}.

Small imperfections in the alignment of the ejection optics with the solenoidal magnetic field result in an ellipse-shaped orbital projection on the PS-MCP. To minimize the effects of these distortions, all reference and final phase measurements are taken within 10$^{\circ}$ of each other. Any temporal instabilities in the magnetic field are accounted for by measuring calibrant isotopes close in time to IOI measurements. 

Additional systematics, such as ion-ion interactions, magnetic field imperfections and electric field instabilities, have been determined to have a resulting uncertainty of $4.0 \times10^{-9}$ \cite{Ray2024}; this has been added in quadrature.

In total, one Pd mass and two Ru masses were measured with the CPT. At each mass unit, IOIs and other beam species were identified by taking data at a series of accumulation times between 5 ms to 100 ms. Once species are identified, measurements of the IOI are taken at a number of accumulation times around $\approx$ 450 ms, within a window of $\approx$ 1 ms. Accumulation times were carefully chosen such that contaminant species spots did not overlap with the Pd or Ru spot. When necessary, data when more than five ions were present in the trap simultaneously were discarded to minimize effects due to ion-ion interactions. Sufficient data was taken to achieve $\approx$ 2 mHz uncertainty on relevant cyclotron frequencies.

\begin{table*}[ht]

  \centering
  \caption{Results of the mass measurements performed, compared to the literature values recommended by AME2020 \cite{Wang2021}. Included is the frequency ratio ($\nu_{c,\text{IOI}}/\nu_{c,\text{ref}}$) between the ionic masses of the ion of interest (IOI) and the reference ion for all measurements. Differences are $m_{\text{new}} - m_{\text{lit}}$. All mass values are in keV. 
  }
    \begin{tabular}{c c c c c c}
    \toprule    Nuclide & Mass Excess & Literature & Difference & $\quad$ Reference Ion & Frequency Ratio \\  \hline \hline

  \\ $^{108}$Ru$^+$ &  -83659.9\,\,(12) & -83661(9) & 1.1\,\,(88) &  	$\quad$  $^{108}$Pd$^+$	 	&  0.999\,941\,658\,355\,\,\,(403)	  \\
 
 $^{110}$Ru$^+$ &  -80072.83(77) & -80073(9) & 0.17(900) &  	$\quad$  $^{110}$Pd$^+$     &  0.999\,919\,342\,130\,\,\,(391)	  \\
 
 $^{116}$Pd$^+$ &  -79833.04(81) & -79831(7) & -2.0\,(72) &  	$\quad$  $^{116}$Cd$^+$ 	&  0.999\,917\,762\,315\,\,\,(738)     \\ \\ 
 
 \hline \hline

    \end{tabular}%
  \label{tab:mass_table}%
\end{table*}

\section{Results}

Table \ref{tab:mass_table} reports the mass excesses of all isotopes measured in this work, as well their mass excesses as found in literature. The suggested literature values from AME2020 are entirely derived from previous measurements at JYFLTRAP at the University of Jyv\"{a}skyl\"{a} \cite{Hager2007}. These measurements were made to an uncertainty of just under 10 keV using the ToF-ICR technique \cite{Dilling2018}. Our data are in good agreement with these earlier results, and represent an improvement in uncertainty by approximately an order of magnitude, demonstrating the increased precision attainable with the PI-ICR technique.

The measured Ru and Pd mass excesses were used in the training of the PIML model, which uses a MDN \cite{Bishop1994} to map input data to an admixture of Gaussian distributions. This effectively accounts for the probabilistic nature of the input experimental data, and results in posterior distributions for the outputs. Previous studies have shown that an expanded feature space provides for better mass predictions and overall model convergence \cite{Lovell2022}. With this in mind, our chosen physics-based feature space consists of nine total features as in \cite{Mumpower2023}: proton number ($Z$), neutron number ($N$), mass number ($A$), measures of the odd-even pairing effects due to protons ($Z_{\text{eo}}$), neutrons ($N_{\text{eo}}$) and all nucleons ($A_{\text{eo}}$), the number of valence protons ($V_p$) and neutrons ($V_n$), and a measure of the proton-neutron isospin asymmetry ($P_\text{asym} = \frac{N-Z}{A}$). All features are functions solely of $N$ and $Z$. Input training data is a hybrid mixture of randomly-selected experimental data from AME2020 \cite{Wang2021} and theoretical model data from FRDM2012 \cite{Moller2016}, WS4 \cite{Wang2014} and DZ33 \cite{Duflo1995}. We employ a network of six hidden layers and ten hidden nodes per layer with an Adam optimizer with learning rate 0.0002 \cite{Kingma2014} and a weight-decay regularization of 0.01. In line with previous studies \cite{Lovell2022}, network complexity was minimized while maintaining good agreement between training set data and resulting model predictions. Training occurs via the minimization of two independent log-likelihood loss functions: $\mathcal{L}_1$ which contains information on the input hybrid mass excess data, and $\mathcal{L}_2$ which help constrain model results to follow the Garvey-Kelson (GK) relations, presented in detail in \cite{Garvey1969}. This additional physics-based constraint has been shown to aid significantly in the improvement of model fitting results. The total loss function is given as:

\begin{equation}
\mathcal{L}_{\text{total}} = \mathcal{L}_1 + \lambda_{\text{phys}}\mathcal{L}_2
\end{equation}

\noindent where $\lambda_{\text{phys}}$ is an additional tunable hyperparameter that controls the constraining power of the GK relations in a given model fit. In line with previous findings in \cite{Mumpower2023}, we use $\lambda_{\text{phys}} = 1.0$. We conclude the training process via a cross-validation method with the experimental mass values, which halts training when the associated cross-validation loss has increased for a long period of time without any decrease in $\mathcal{L}_{\text{total}}$.

\begin{figure}[tb]
    \begin{center}
        \includegraphics[width=0.925\columnwidth]{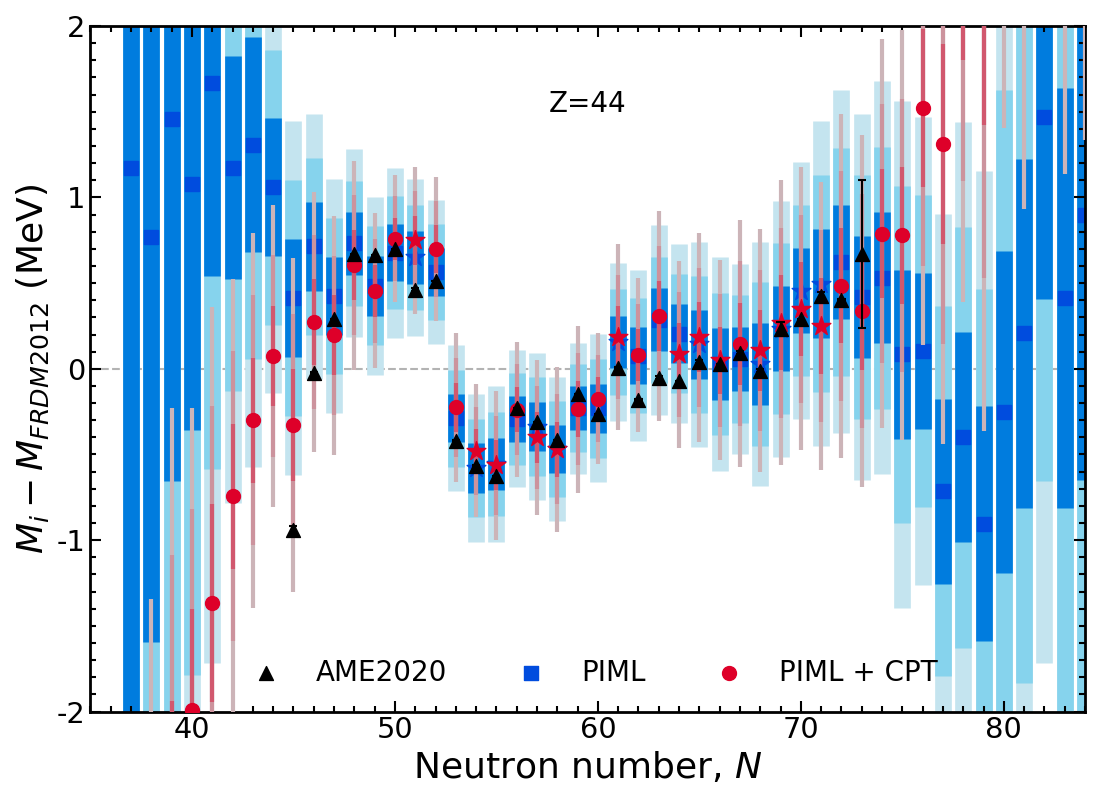}
        \caption{A plot comparing two different PIML model mass results and mass values from AME2020\cite{Wang2021} for Ru isotopes ($Z = 44$). Blue squares represent the PIML model and red circles the PIML$_\text{CPT}$ model, with the different shaded bands representing the 1$\sigma$, 2$\sigma$ and 3$\sigma$ uncertainties. Black triangles and error bars represent mass values and 1$\sigma$ uncertainties from \cite{Wang2021}. Stars indicate masses that were used as part of the training dataset.} 
        \label{fig:piml_ME}
    \end{center}
\end{figure}

Mass excess results for Ru masses from two different PIML models are presented alongside experimental values as reported in AME2020 in Fig. \ref{fig:piml_ME}. Each model is trained with the same hybrid set of randomly selected masses from AME2020 \cite{Wang2021}, the masses of $^{108}$Ru, $^{110}$Ru and $^{116}$Pd, and the three employed mass models with one difference; the PIML$_\text{CPT}$ model uses the three mass values for $^{108}$Ru, $^{110}$Ru, and $^{116}$Pd from this work while the PIML model uses the mass values from AME2020. This technique, a Bayesian update of a ML model with new information, enables unique studies of the effects of individual mass results on both local and global model trends. Both models provide an excellent fit globally to AME mass data, with $\sigma_{\text{RMS},\text{PIML}} = 0.380$ MeV and $\sigma_{RMS,\text{PIML-CPT}} = 0.324$ MeV. A notable difference in the mass trends is apparent between the two models for the Ru masses; while both models do similarly well in modeling masses closer to stability, the PIML$_\text{CPT}$ model appears to match more closely the trends of the neutron-deficient end of the isotope chain. A similar such behavior is seen in the Pd masses.

\begin{figure}[tb]
    \begin{center}
        \includegraphics[width=1\columnwidth]{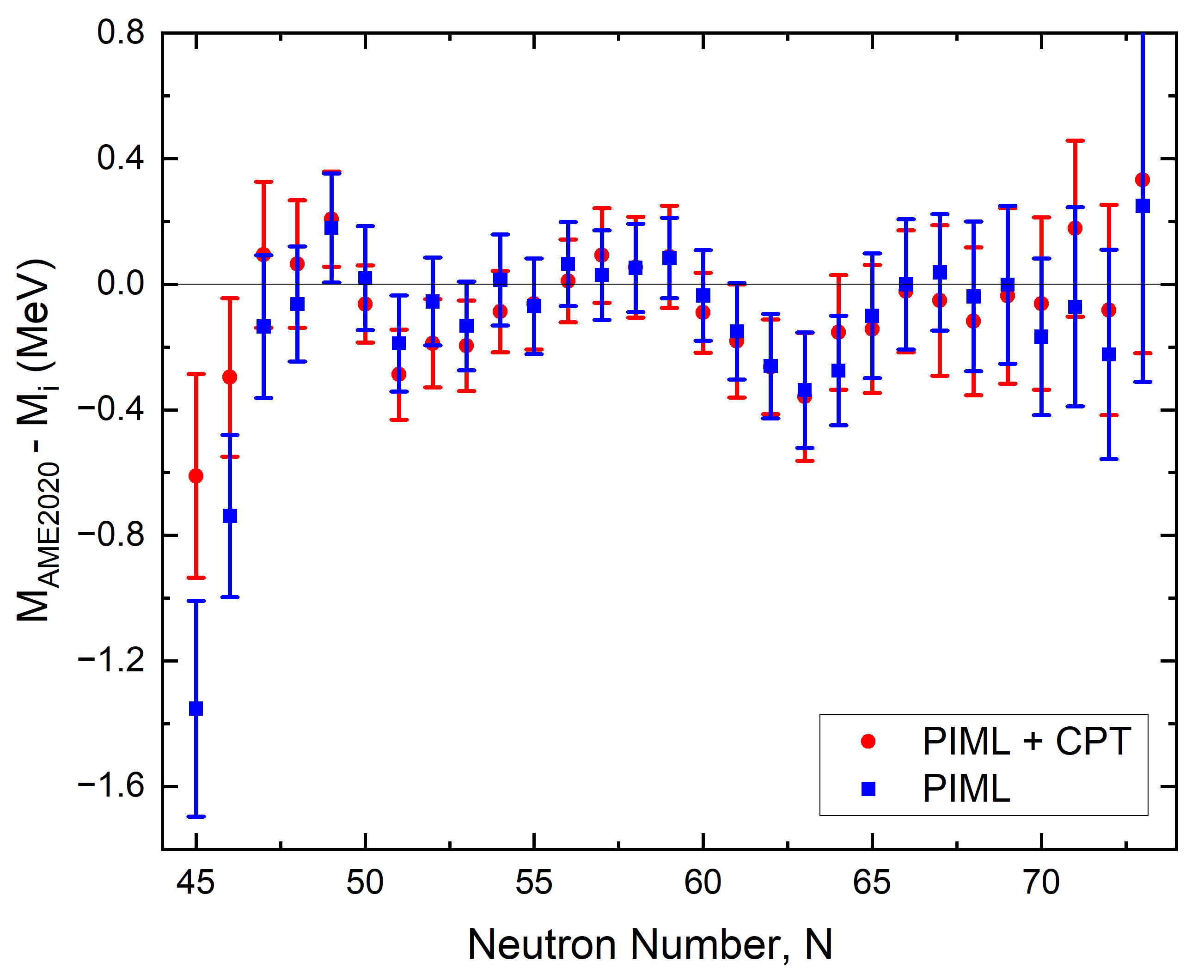}
        \caption{A plot showing the difference between two different PIML model mass results and mass values from AME2020\cite{Wang2021} for the Ru ($Z = 44$) isotopes. Blue squares represent the PIML model and red circles the PIML$_\text{CPT}$ model. Error bars represent the 1$\sigma$ uncertainties from the respective PIML model.} 
        \label{fig:piml_diff}
    \end{center}
\end{figure}

An alternative visualization of the results from the two models is presented in Fig. \ref{fig:piml_diff}, which shows the difference between each of the models' results and AME2020 \cite{Wang2021} for the Ru isotopes. Of note, the mass prediction for $^{108}$Ru ($N = 64$) is improved for the PIML$_\text{CPT}$ model such that the AME2020 mass and PIML$_\text{CPT}$ mass are consistent within 1$\sigma$. There is no statistically significant improvement for $^{110}$Ru ($N = 66$), as the PIML model already well-predicted the AME mass.

These calculations highlight the importance of well-measured masses and the significance they can have on the extrapolative behavior of global ML mass models. Further mass measurements, both more exotic and precise, will enable even further constraint of global nuclear mass models. 

\section{Summary}

Measurements of neutron-rich Ru and Pd isotopes were performed at the CARIBU facility at ANL using the Canadian Penning Trap. All mass measurements represent an improvement in precision of approximately an order of magnitude over earlier results, demonstrating the precision capabilities of the PI-ICR technique over previously existing techniques. These data were used in the training of PIML models, which enables mass excess modeling via a mixture density neural network. The results demonstrate the significant effect a small number of precise mass data can have on resulting trends in the mass surface. Further precise mass measurements, alongside developments in machine learning approaches, are ultimately needed to improve our nuclear mass interpolative and predictive capabilities.

\begin{acknowledgments}

This work was performed with the support of US Department of Energy, Office of Nuclear Physics under Contract No. DE-AC02-06CH11357 (ANL), the Natural Sciences and Engineering Research Council of Canada under Grant No. SAPPJ-2018-00028, and the US National Science Foundation under Grant No. PHY-2011890 and PHY-2310059, and under the auspices of the U.S. Department of Energy by Lawrence Livermore National Laboratory under Contract No. DE-AC52-07NA27344. This research used resources of ANL’s ATLAS facility, which is a DOE Office of Science User Facility. MRM acknowledges support from the Directed Asymmetric Network Graphs for Research (DANGR) initiative at Los Alamos National Laboratory (LANL). LANL is operated by Triad National Security, LLC, for the National Nuclear Security Administration of U.S. Department of Energy (Contract No. 89233218CNA000001). ML acknowledges support from NSF Grant No. PHY-2020275 [Network for Neutrinos, Nuclear Astrophysics and Symmetries (N3AS)]. WSP, DR, AAV, MB and DEMH acknowledge the contributions of IOB and BHB.

\end{acknowledgments}

\bibliography{library}
    
\end{document}